\documentstyle[pra,aps,twocolumn,epsfig,floats]{revtex}    

\newcommand{\tfrac}[2]{{\textstyle\frac{#1}{#2}}}
\newcommand{\ETFD}{E^{\mathrm{(TFD)}}}
\newcommand{\Ekin}{E_{\mathrm{kin}}}
\newcommand{\Etrap}{E_{\mathrm{trap}}}
\newcommand{\Edd}{E_{\mathrm{dd}}}
\newcommand{\Vdd}{V_{\mathrm{dd}}}
\newcommand{\bVdd}{\overline{V}_{\mathrm{dd}}}
\newcommand{\thalf}{{\textstyle\frac{1}{2}}}
\newcommand{\half}{\frac{1}{2}}
\newcommand{\third}{\frac{1}{3}}
\newcommand{\fifth}{\frac{1}{5}}
\newcommand{\sixth}{\frac{1}{6}}
\newcommand{\D}{{\mathrm{d}}}
\newcommand{\I}{{\mathrm{i}}}
\newcommand{\svec}[1]{\!\vec{\,#1}}
\newcommand{\Exp}[1]{\,\mathrm{e}\power{#1}}%
\newcommand{\power}[1]{^{\mbox{\footnotesize$#1$}}}

\begin{document}

\draft

\title{Semiclassical theory of trapped fermionic dipoles}

\author{Krzysztof G\'{o}ral,$^{1}$ Berthold-Georg Englert,$^{2,3}$ and
Kazimierz Rz\c{a}\.{z}ewski$^{1}$}
\address{%
$^1$Center for Theoretical Physics and College of Science,\\
Polish Academy of Sciences, Aleja Lotnik\'ow 32/46, 02-668 Warsaw, Poland\\
$^2$Max-Planck-Institut f\"ur Quantenoptik,
Hans-Kopfermann-Strasse 1, 85748 Garching, Germany \\
$^3$Atominstitut, Technische Universit\"at Wien,
Stadionallee 2, 1020 Wien, Austria}

\maketitle

\begin{abstract}\hspace*{1em}%
We investigate the properties of a degenerate dilute gas of neutral
fermionic particles in a harmonic trap that interact via dipole-dipole forces.
We employ the semiclassical Thomas-Fermi method and discuss the
Dirac correction to the interaction energy.
A nearly analytic as well as an exact numerical minimization of the
Thomas-Fermi-Dirac energy functional are performed in order to obtain
the density distribution.
We determine the stability of the system as a function of the
interaction strength, the particle number, and the trap geometry.
We find that there are interaction strengths and particle numbers for which
the gas cannot be trapped stably in a spherically symmetric trap, but
both prolate and oblate traps will work successfully.
\end{abstract}


\narrowtext

\section{Introduction}
The experimental achievement of quantum degeneracy in a dilute trapped
gas of cold fermionic atoms \cite{JILAfermions} has stimulated
theoretical interest in the properties of this fundamental system.
Attention has  focused on the critical temperature
\cite{pairingTc} and the detection \cite{BCSdetection} of Cooper
pairing as well as on the properties of mixtures of various
fermionic and bosonic species \cite{mixtures,phonon}.
Another important problem concerns the interactions between ultracold
fermions \cite{phonon,Rokhsar,interFermi}.
Owing to the exclusion principle,
spin-polarized Fermi atoms do not interact via s-wave collisions,
whereas they dominate in the low-energy regime for bosons and have
pronounced effects on the statics and dynamical properties of cold
boson gases \cite{RMP}.
Hence, in the absence of low-energy collisions other types of
forces come into play.

A good candidate is a dipole-dipole interaction between atoms or molecules,
not analyzed so far in the context of cold trapped fermions.
Some atoms possess permanent magnetic dipole moments of considerable
magnitude (chromium, for instance, has $\mu=6 \mu_B$).
It was also proposed to induce electric dipoles in atoms
\cite{Marinescu,Santos}.
Huge permanent electric dipole moments occur naturally in diatomic polar
molecules \cite{polarmol}.
The behavior of atomic bosonic dipoles in traps has been investigated
in \cite{You,Goral}, which addressed the question of instabilities
in the system caused by an attractive component of dipolar interactions.
The conclusion drawn was that a large enough positive scattering
length (providing repulsive interactions) can stabilize a system
of bosonic dipoles.
For strong (e.g. molecular) dipoles, when supposedly the scattering length
can be neglected, it is the trap geometry that plays a crucial
role \cite{Santos} -- the system is stable provided the trap
assures the domination of repulsive interactions in the gas.

A full quantum mechanical description of the system of many
interacting fermions is of course very complex.
But the lessons of semiclassical atomic physics can be applied,
in particular the Thomas-Fermi approach \cite{Thomas26,Fermi27} and its
refinements (see, e.\,g., \cite{LNP300}).
Its success in describing static properties of atoms is well known,
and we note that, in recent years, these methods were also used
successfully for studying dynamical processes of atoms and molecules
in superstrong light pulses \cite{superstrong}.

Our paper is organized as follows: in Section II, the Thomas-Fermi
model is revisited with an eye on the dipole-dipole forces. The
Dirac correction to the interaction energy term is discussed and
scaling properties derived with the help of the virial theorem. In
Section III the results of approximate (nearly analytic) and
numerical minimizations of the Thomas-Fermi-Dirac energy functional are
presented. Unexpectedly, we find that the system may withstand
larger dipolar forces both for very flat and for highly elongated
traps.

\section{Thomas-Fermi model}

\subsection{General considerations}

Let's begin by recalling some basic things, mainly collected from
Refs.\ \cite{LNP300,BGE92,MC+BGE93}, with due attention to the
changed situation: here fully spin-polarized fermions --- there
fermions with no net spin. The spatial one-particle  density is
denoted by $n(\svec{r})$, it is normalized to the total particle
number $N$,
\begin{equation}
  \label{eq:N}
  N=\int(\D\svec{r})\,n(\svec{r})\,,
\end{equation}
where $(\D\svec{r})\equiv\D x\D y\D z$ denotes the volume element.
The spatial one-particle density matrix
$n^{(1)}(\svec{r}';\svec{r}'')$ and the one-particle Wigner
function $\nu(\svec{r},\svec{p})$ are related by
\begin{equation}
  \label{eq:Wig1}
  n^{(1)}(\svec{r}';\svec{r}'')=\!\int\!\frac{(\D\svec{p})}{(2\pi\hbar)^3}
\nu\bigl(\thalf(\svec{r}'+\svec{r}''),\svec{p}\bigr)
\Exp{\I\svec{p}\cdot(\svec{r}'-\svec{r}'')/\hbar}\;.
\end{equation}
The spatial and momental one-particle densities are obtained by
integrating $\nu(\svec{r},\svec{p})$ over the other variable,
\begin{eqnarray}
 n(\svec{r})= n^{(1)}(\svec{r};\svec{r})
            &=&\int\frac{(\D\svec{p})}{(2\pi\hbar)^3}\nu(\svec{r},\svec{p})\,,
\nonumber\\
\rho(\svec{p})
& =&\int\frac{(\D\svec{r})}{(2\pi\hbar)^3}\nu(\svec{r},\svec{p})\,.
\label{eq:Wig2}
\end{eqnarray}
They are needed for the calculation of the kinetic energy,
\begin{equation}
  \label{eq:Ekin1}
  \Ekin=\int(\D\svec{p})\,\frac{\svec{p}^2}{2M}\rho(\svec{p})\,,
\end{equation}
and the external potential energy (of the harmonic trap),
\begin{equation}
  \label{eq:Etrap1}
  \Etrap=\int(\D\svec{r})\,\thalf M \omega^2\bigl[x^2+y^2+(\beta z)^2\bigr]
n(\svec{r})\,,
\end{equation}
where $M$ is the mass of the atom species considered, $\omega$ is the
(transverse) trap frequency, and $\beta$ is the aspect ratio of the
cylindrical trap.
The trap is spherically symmetric for $\beta=1$;
for $\beta<1$, the equipotential surfaces are prolate (``cigar shaped'')
ellipsoids;
for $\beta>1$, they are oblate (``lentil shaped'') ellipsoids.

For the dipole-dipole interaction energy, $\Edd$, we need (the
diagonal part of) the two-particle density matrix
$n^{(2)}(\svec{r}_1',\svec{r}_2';\svec{r}_1'',\svec{r}_2'')$,
\begin{equation}
  \label{eq:Edd1}
  \Edd=\half\int(\D\svec{r}')(\D\svec{r}'')\Vdd(\svec{r}'-\svec{r}'')
       n^{(2)}(\svec{r}',\svec{r}'';\svec{r}',\svec{r}'')\,,
\end{equation}
where
\begin{equation}
  \label{eq:Vdd1}
  \Vdd(\svec{r})=\frac{\mu_0}{4\pi}\left[
\frac{\svec{\mu}^2}{r^3}-3\frac{(\svec{\mu}\cdot\svec{r})^2}{r^5}
-\frac{8\pi}{3}\svec{\mu}^2\delta(\svec{r})\right]\,.
\end{equation}
We note that the contact term, proportional to $\delta(\svec{r})$,
is required by the condition that the magnetic field made by the point
dipole be divergence-free \cite{comment}.
An alternative way of presenting $\Vdd$ is
\begin{equation}
  \label{eq:Vdd2}
  \Vdd(\svec{r})=\frac{\mu_0}{4\pi}\svec{\mu}\cdot\left[
                 -\svec{\nabla}\,\svec{\nabla}\frac{1}{r}
                 -4\pi\tensor{1}\delta(\svec{r})\right]\cdot\svec{\mu}\,,
\end{equation}
which is a particularly convenient starting point for evaluating
the Fourier transform
\begin{equation}
  \label{eq:Vdd3}
  \int(\D\svec{r})\,\Exp{\I \svec{k}\cdot\svec{r}}\Vdd(\svec{r})=
\frac{\mu_0}{4\pi}\svec{\mu}\cdot\left[4\pi\frac{\svec{k}\,\svec{k}}{k^2}
                                       -4\pi\tensor{1}\right]\cdot\svec{\mu}\,.
\end{equation}
The vanishing divergence just mentioned is here immediately
recognized, inasmuch as $\svec{k}\cdot\bigl[\cdots\bigr]=0$.

\subsection{Thomas-Fermi-Dirac functionals}
The semiclassical approximation now employed --- in the spirit of
what the TFD trio (Thomas \cite{Thomas26}, Fermi \cite{Fermi27},
and Dirac \cite{Dirac30}) did, although
in a technically different manner --- is two-fold: We approximate
$n^{(2)}$ by products of $n^{(1)}$ factors (Dirac),
\begin{eqnarray}
  n^{(2)}(\svec{r}_1',\svec{r}_2';\svec{r}_1'',\svec{r}_2'')
&=&n^{(1)}(\svec{r}_1';\svec{r}_1'')n^{(1)}(\svec{r}_2';\svec{r}_2'')
\nonumber\\&&
\mbox{}-n^{(1)}(\svec{r}_1';\svec{r}_2'')n^{(1)}(\svec{r}_2';\svec{r}_1'')\,,
  \label{eq:n2TFD}
\end{eqnarray}
and $n^{(1)}$ by a brutally simple Wigner function
(Thomas and Fermi),
\begin{equation}
  \label{eq:WigTF}
  \nu(\svec{r},\svec{p})
=\eta\bigl(\hbar^2[6\pi^2n(\svec{r})]^{\frac{2}{3}}-\svec{p}^2\bigr)\,,
\end{equation}
where $\eta(\ )$ is Heaviside's unit step function. This gives the
density functional of the kinetic energy as
\begin{equation}
  \label{eq:EkinTFD}
  \Ekin[n]=\int(\D\svec{r})\,\frac{\hbar^2}{M}\frac{1}{20\pi^2}
\bigl[6\pi^2n(\svec{r})\bigr]^{\frac{5}{3}}\,,
\end{equation}
and the density functional for the potential energy in the trap is
$\Etrap$ of (\ref{eq:Etrap1}).

The dipole-dipole interaction energy consists of two parts,
corresponding to the two summands in (\ref{eq:n2TFD}),
\begin{equation}
  \label{eq:Edd2}
  \Edd[n]=\Edd^{\mathrm{(dir)}}[n]+\Edd^{\mathrm{(ex)}}[n]
\end{equation}
with the direct term
\begin{equation}
  \label{eq:Edd3}
  \Edd^{\mathrm{(dir)}}[n]=\half\int(\D\svec{r}')(\D\svec{r}'')
                  n(\svec{r}')\Vdd(\svec{r}'-\svec{r}'')n(\svec{r}'')
\end{equation}
and the exchange term
\begin{eqnarray}
  \Edd^{\mathrm{(ex)}}[n]&=&-\half\int(\D\svec{r}')(\D\svec{r}'')\,
    n^{(1)}(\svec{r}';\svec{r}'')n^{(1)}(\svec{r}'';\svec{r}')
   \nonumber\\
&&\hphantom{-\half\int(\D\svec{r}')(\D\svec{r}'')}\quad\times
 \Vdd(\svec{r}'-\svec{r}'')
\nonumber\\
&=&-\half\int(\D\svec{r})\int(\D\svec{s})\,\Vdd(\svec{s})\nonumber\\
&&\times
  n^{(1)}(\svec{r}+\thalf\svec{s};\svec{r}-\thalf\svec{s})
  n^{(1)}(\svec{r}-\thalf\svec{s};\svec{r}+\thalf\svec{s})\,.
\nonumber\\
\label{eq:Edd4}
\end{eqnarray}
Now we note that
\begin{eqnarray}
 &&n^{(1)}(\svec{r}+\thalf\svec{s};\svec{r}-\thalf\svec{s})
   n^{(1)}(\svec{r}-\thalf\svec{s};\svec{r}+\thalf\svec{s})
\nonumber\\&&\hspace*{5em}=
\int\frac{(\D\svec{p}')(\D\svec{p}'')}{(2\pi\hbar)^6}
\nu(\svec{r},\svec{p}')  \nu(\svec{r},\svec{p}'')
\Exp{\I(\svec{p}'-\svec{p}'')\cdot\svec{s}/\hbar}
\nonumber\\  \label{eq:Edd5}
\end{eqnarray}
depends only on the length $s=\bigl|\svec{s}\bigr|$ of vector
$\svec{s}$, not on its direction $\svec{s}/s$, because --- in the
TF approximation (\ref{eq:WigTF}) --- the product
$\nu(\svec{r},\svec{p}')\nu(\svec{r},\svec{p}'')$ involves only
$p'=\bigl|\svec{p}'\bigr|$ and $p''=\bigl|\svec{p}''\bigr|$. As a
consequence, it is permissible to replace, in (\ref{eq:Edd4}),
$\Vdd(\svec{s})$ by its average over the solid angle associated
with $\svec{s}$,
\begin{equation}
  \label{eq:Vdd4}
  \Vdd(\svec{s})\to\frac{\mu_0}{4\pi}\left[
-\frac{8\pi}{3}\svec{\mu}^2\delta(\svec{s})\right]\,.
\end{equation}
We thus arrive at
\begin{eqnarray}
\Edd^{\mathrm{(ex)}}[n]&=&\half\int(\D\svec{r})\left[n(\svec{r})\right]^2
\,\frac{\mu_0}{4\pi}\,\frac{8\pi}{3}\,\svec{\mu}^2 \nonumber\\
&=&\half\int(\D\svec{r})(\D\svec{r}')n(\svec{r})
\frac{\mu_0}{4\pi}\left[
\frac{8\pi}{3}\svec{\mu}^2\delta(\svec{r}-\svec{r}')\right]n(\svec{r}')\,,
\nonumber\\   \label{eq:Edd6}
\end{eqnarray}
and accordingly
\begin{equation}
  \label{eq:Edd7}
\Edd[n]=\half\int(\D\svec{r})(\D\svec{r}')\,
n(\svec{r})\bVdd(\svec{r}-\svec{r}')n(\svec{r}')
\end{equation}
with
\begin{equation}
  \label{eq:bVdd}
 \bVdd(\svec{r})= \frac{\mu_0}{4\pi}\left[
\frac{\svec{\mu}^2}{r^3}-3\frac{(\svec{\mu}\cdot\svec{r})^2}{r^5}
\right]\,,
\end{equation}
which is $\Vdd(\svec{r})$ of (\ref{eq:Vdd1}) with the contact term removed.
In (\ref{eq:Vdd2}) and (\ref{eq:Vdd3}) this corresponds to multiplying the unit
dyadic $\tensor{1}$ by $\third$ \cite{Goral}.

The rotational symmetry of the trap potential in (\ref{eq:Etrap1})
distinguishes the $z$ axis, and we take for granted that this is also the
direction of spin polarization,
\begin{equation}
  \svec{\mu}=\mu\svec{e}_z\;.
\end{equation}
Then
\begin{equation}
  \label{eq:bVdd'}
  \bVdd(\svec{r})= \frac{\mu_0}{4\pi}\mu^2\frac{1-3(z/r)^2}{r^3}\,,
\end{equation}
and the whole system is invariant under rotations around the $z$ axis.

The total TFD energy functional is then given by the sum of the
kinetic energy (\ref{eq:EkinTFD}), the potential energy in the
trap (\ref{eq:Etrap1}), and the dipole-dipole interaction energy
(\ref{eq:Edd7}),
\begin{eqnarray}
\ETFD[n] &=&\!\int\!(\D\svec{r})
\biggl[\frac{\hbar^2}{M}\frac{1}{20\pi^2}
\bigl[6\pi^2n(\svec{r})\bigr]^{\frac{5}{3}} +\thalf
M\omega^2r^2n(\svec{r})\biggr] \nonumber\\ &&
+\half\int(\D\svec{r})(\D\svec{r}')n(\svec{r})\bVdd(\svec{r}-\svec{r}')
n(\svec{r}')\,.
  \label{eq:EtotTFD}
\end{eqnarray}
The density that minimizes $\ETFD$ under the constraint
(\ref{eq:N}) obeys the nonlinear integral equation
\begin{eqnarray}
&&  \frac{\hbar^2}{2M}\bigl[6\pi^2n(\svec{r})\bigr]^{\frac{2}{3}}
+\thalf M\omega^2\bigl[x^2+y^2+(\beta z)^2\bigr]
\nonumber\\&&\qquad\qquad+\int(\D\svec{r}')\bVdd(\svec{r}-\svec{r}')
n(\svec{r}') =\thalf M\omega^2R^2\,,
  \label{eq:NonLinIntEq}
\end{eqnarray}
where $\thalf M\omega^2R^2$ is a convenient way of writing the
Lagrange multiplier for the constraint.

\subsection{Virial theorems}\label{sec:virial}
Scaling transformations of the form
\begin{equation}
  \label{eq:scale1}
  n(\svec{r})\to\lambda^{3+\alpha}n(\lambda\svec{r})\,,\quad
N\to\lambda^{\alpha}N
\end{equation}
are consistent with the constraint (\ref{eq:N}). They affect the
various pieces of $\ETFD$ in accordance with
\begin{eqnarray}
\Ekin&\to&\lambda^{2+\frac{5}{3}\alpha}\Ekin\,,\nonumber\\
\Etrap&\to&\lambda^{-2+\alpha}\Etrap\,,\nonumber\\
\Edd&\to&\lambda^{3+2\alpha}\Edd\,,
  \label{eq:scale2}
\end{eqnarray}
so that ($E\equiv\ETFD$)
\begin{eqnarray}
  E&=&\Ekin+\Etrap+\Edd\nonumber\\
&\to&\lambda^{2+\frac{5}{3}\alpha}\Ekin
+\lambda^{-2+\alpha}\Etrap+\lambda^{3+2\alpha}\Edd\,.
  \label{eq:scale3}
\end{eqnarray}
In the infinitesimal vicinity of $\lambda=1$, all first-order
changes of $E$ originate in the explicit change of $N$, $\delta
N=\delta\lambda\,\alpha N$, and therefore
\begin{equation}
  \label{eq:scale4}
  \alpha N\frac{\partial E}{\partial N}=(2+\tfrac{5}{3}\alpha)\Ekin
+(-2+\alpha)\Etrap+(3+2\alpha)\Edd
\end{equation}
must hold irrespective of the value of parameter $\alpha$. In view
of the linear $\alpha$ dependence, we get two independent
statements,
\begin{eqnarray}
\alpha=0\,:&\quad&2\Ekin-2\Etrap+3\Edd=0\,,\nonumber\\
\alpha=-\tfrac{3}{2}\,:&&\Ekin+7\Etrap=3 N\frac{\partial
E}{\partial N}\,.
  \label{eq:scale5}
\end{eqnarray}
They enable us to express $\Ekin$, $\Etrap$, and $\Edd$ in terms
of $E$ and $N\partial E/\partial N$,
\begin{eqnarray}
\Ekin&=&\frac{21}{2}E-\frac{15}{2}N\frac{\partial E}{\partial
N}\,, \nonumber\\
\Etrap&=&-\frac{3}{2}E+\frac{3}{2}N\frac{\partial E}{\partial
N}\,, \nonumber\\ \Edd&=&-8E+6N\frac{\partial E}{\partial N}\,.
  \label{eq:scale6}
\end{eqnarray}
In conjunction with ($\mu=\bigl|\svec{\mu}\bigr|$)
\begin{eqnarray}
\mu\frac{\partial}{\partial\mu}E&=&2\Edd\,,\nonumber\\
\omega\frac{\partial}{\partial\omega}E&=&2\Etrap\,,\nonumber\\
M\frac{\partial}{\partial M}E&=&\Etrap-\Ekin\,,
  \label{eq:scale7}
\end{eqnarray}
they imply that the ground state energy $E(M,\omega,\mu,N)$ is of
the form (we leave the $\beta$ dependence implicit)
\begin{equation}
  \label{eq:scale8a}
 E(M,\omega,\mu,N)=\hbar\omega N^{\frac{4}{3}}e(N^{\frac{1}{6}}\varepsilon)
\end{equation}
with
\begin{equation}
  \label{eq:scale8b}
 \varepsilon=\left(\omega
M^3/\hbar^5\right)^{\half}\frac{\mu_0}{4\pi}\mu^2 \,.
\end{equation}
Clearly, the dimensionless number $\varepsilon$ measures the
relative strength of the dipole-dipole  interaction. The universal
function $e(\ )$ is to be found numerically
(for each $\beta$ value of interest).
For $\varepsilon=0$,
we can solve (\ref{eq:NonLinIntEq}) immediately,
\begin{equation}
  \label{eq:n0}
  n(\svec{r})=\frac{1}{6\pi^2}\bigl(M\omega/\hbar\bigr)^3
              \bigl[R^2-x^2-y^2-(\beta z)^2\bigr]^{\frac{3}{2}}\,,
\end{equation}
where, by convention, $[\cdots]^{\frac{3}{2}}=0$ for negative arguments, and
\begin{equation}
  \label{eq:R0}
  R=\bigl(48\beta N\bigr)^\sixth\bigl(M\omega/\hbar\bigr)^{-\half}
\end{equation}
as a consequence of (\ref{eq:N}), and so we find
\begin{equation}
  \label{eq:e0}
e_0\equiv e(0)=\frac{3}{4}(6\beta)^\third=1.363\beta^\third\,.
\end{equation}
For the spherically symmetric case $\beta=1$,
$\Edd$ vanishes in first-order perturbation theory, so that
\begin{equation}
  \label{eq:scale9}
\beta=1\,:\quad
  e(N^{\frac{1}{6}}\varepsilon)=2^{-\frac{5}{3}}3^{\frac{4}{3}}
                                +e_2 N^{\frac{1}{3}}\varepsilon^2
                                +O(N^{\frac{1}{2}}\varepsilon^3)
\end{equation}
with $e_2<0$.

\subsection{Dimensionless variables}
These scaling laws invite the use of correspondingly chosen dimensionless
variables, such as
\begin{equation}
  \label{eq:xvec}
  \svec{x}=\svec{r}/a\quad\text{with}\enskip
        a=N^{\frac{1}{6}}\sqrt{\frac{\hbar}{M\omega}}
\end{equation}
for the position and
\begin{equation}
  \label{eq:gdef}
  g(\svec{x})=\frac{a^3}{N}n(a\svec{x})\quad\text{or}\quad
  n(\svec{r})=\frac{N}{a^3}g(\svec{r}/a)
\end{equation}
for the density.
The constraint (\ref{eq:N}) then appears as
\begin{equation}
  \label{eq:N'}
  \int(\D\svec{x})\,g(\svec{x})=1\,,
\end{equation}
and the TFD energy acquires the form
\begin{eqnarray}
  \frac{\ETFD}{\hbar\omega N^{\frac{4}{3}}}[g]&=&
\frac{3}{10}(6\pi^2)^{\frac{2}{3}} \int(\D\svec{x})\,
[g(\svec{x})]^{\frac{5}{3}} \nonumber \\ &&+\frac{1}{2} \int
(\D\svec{x})\, \bigl(x_1^2+x_2^2+\beta^2x_3^2\bigr)g(\svec{x}) \nonumber\\
&&+\half N^{\frac{1}{6}}\varepsilon
\int (\D\svec{x})(\D\svec{x}')\,g(\svec{x})
\frac{1-3\cos^{2}\theta}{|\svec{x}-\svec{x}'|^3}g(\svec{x}')\,,
\nonumber\\  \label{eq:scaledETFD}
\end{eqnarray}
where $\theta$ is the angle between the polarization direction (the
$x_3$ direction) and the relative position vector $\svec{x}-\svec{x}'$,
\begin{equation}
  \label{eq:theta}
  (\svec{x}-\svec{x}')\cdot\svec{\mu}
=\bigl|\svec{x}-\svec{x}'\bigr|\mu\cos\theta
=(x_3-x_3')\mu\,.
\end{equation}
The minimum of the scaled TFD energy is $e(N^{\frac{1}{6}}\varepsilon)$ of
(\ref{eq:scale8a}) (with its implicit $\beta$ dependence); it is obtained
for the
$g(\svec{x})$ that obeys the dimensionless analog of (\ref{eq:NonLinIntEq}),
\begin{eqnarray}
&&  \thalf\bigl[6\pi^2g(\svec{x})\bigr]^{\frac{2}{3}}
+\thalf\bigl(x_1^2+x_2^2+\beta^2x_3^2\bigr)
\nonumber\\&&\qquad\qquad+N^{\frac{1}{6}}\varepsilon\int(\D\svec{x}')
\frac{1-3\cos^{2}\theta}{|\svec{x}-\svec{x}'|^3}g(\svec{x}') =\thalf X^2\,,
  \label{eq:NonLinIntEq'}
\end{eqnarray}
where the value of $X=R/a$ is determined by the constraint (\ref{eq:N'}).

\section{Results}
Owing to its intrinsic semiclassical approximations, one expects a few-percent
deviation of the TFD energy from the true ground-state energy.
It is, therefore, not really necessary to solve the nonlinear integral
equation (\ref{eq:NonLinIntEq'}).
A reasonable variational estimate, in conjunction with a few full-blown
numerical solutions for comparison, will do.

\subsection{Gaussian variational ansatz}
The Gaussian ansatz
\begin{equation}
  \label{eq:gauss}
  g(\svec{x})=(2\pi)^{-\frac{3}{2}}\kappa^3\gamma
              \Exp{-\half\kappa^2(x_1^2+x_2^2+\gamma^2x_3^2)}
\end{equation}
is convenient.
In addition to its aspect ratio $\gamma$, the shape parameter,
it contains the scale parameter $\kappa$, so that the virial theorems
of Sec.\ \ref{sec:virial} will be obeyed
for the optimal choice of $\kappa$.
For this scaled density, the scaled kinetic energy is given by
\begin{equation}
  \label{eq:gauss-kin}
  \frac{\Ekin}{\hbar\omega N^{\frac{4}{3}}}=
2^{-\frac{4}{3}} 3^{\frac{19}{6}} 5^{-\frac{5}{2}} \pi^\third
\kappa^2\gamma^{\frac{2}{3}}\,,
\end{equation}
and the scaled value of $\Etrap$ is
\begin{equation}
  \label{eq:gauss-trap}
  \frac{\Etrap}{\hbar\omega N^{\frac{4}{3}}}
=\frac{1}{\kappa^2}\left(1+\frac{\beta^2}{2\gamma^2}\right)\,.
\end{equation}
They exhibit the anticipated dependence on the scale parameter $\kappa$,
and so does the dipole-dipole interaction energy,
\begin{equation}
  \label{eq:gauss-dd}
  \frac{\Edd}{\hbar\omega N^{\frac{4}{3}}}
=  \frac{\kappa^3}{4\sqrt{\pi}}N^\sixth\varepsilon\gamma(\gamma^2-1)
\int_0^1\D\zeta\,\frac{\zeta^2-\zeta^4}{1-\zeta^2+\gamma^2\zeta^2}\;.
\end{equation}
This integral can, of course, be evaluated in terms of elementary functions,
but as it stands we see immediately that the integrand, and thus the integral,
is positive for $0<\gamma<\infty$, so that
\begin{equation}
  \label{eq:gauss-shape}
  \begin{array}[b]{r@{\enskip\mbox{for}\enskip}l@{\quad}l}
  \Edd>0 & \gamma>1 &\text{(oblate Gaussian)},\\[1ex]
  \Edd=0 & \gamma=1 &\text{(spherical Gaussian)},\\[1ex]
  \Edd<0 & \gamma<1 &\text{(prolate Gaussian)},
     \end{array}
\end{equation}
consistent with the expectation that an oblate density has a larger magnetic
interaction energy than a prolate density.
More explicitly, then, we have
\begin{eqnarray}
    \frac{\Edd}{\hbar\omega N^{\frac{4}{3}}}
&=&  \frac{\kappa^3}{4\sqrt{\pi}}N^\sixth\varepsilon\gamma
\nonumber\\[1ex]&&\times\left\{
  \begin{array}{c@{\enskip\mbox{for}\enskip}l@{}}
\displaystyle\!
\frac{1-\vartheta\cot\vartheta}{\sin^2\vartheta}-\third &\displaystyle
\gamma=\frac{1}{\cos\vartheta}>1\,,\\[3ex]
\displaystyle\!
\frac{\vartheta\coth\vartheta-1}{\sinh^2\vartheta}-\third &\displaystyle
\gamma=\frac{1}{\cosh\vartheta}<1\,.
  \end{array}\right.
\nonumber\\  \label{eq:gauss-dd'}
\end{eqnarray}

The Gaussian density (\ref{eq:gauss}) cannot mimic the ${\varepsilon=0}$
solution (\ref{eq:n0}) very well.
But nevertheless, the resulting estimate of the ${\varepsilon=0}$ energy,
\begin{equation}
  \label{eq:gauss-e0}
  \text{Gaussian:}\quad
  e_0=2^{-\sixth}3^{\frac{25}{12}}5^{-\frac{5}{4}}\pi^\sixth\beta^\third
     =1.42\beta^\third
\end{equation}
is only $4.4\%$ in excess of the correct value (\ref{eq:e0}).
This accuracy is sufficient for our purposes; and, in any case, an error of a
few percent is a small price for the enormous simplification that the Gaussian
ansatz brings about.

\begin{figure}[!t]
\begin{center}
\begin{picture}(235,160)
\put(0,-5){\epsfig{file=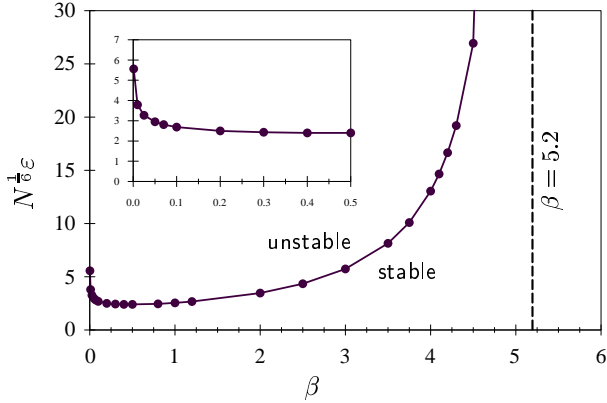}}
\end{picture}
\caption{\label{fig:stability}%
Stability diagram.
The dots, connected by a solid line to guide the eye, define the border
between system parameters ($\beta$: aspect ratio of the trap;
$N^\sixth\varepsilon$: effective interaction strength) for which
$\ETFD$ is bounded from below (stable) or not bounded (unstable).
The inset shows a blow-up of the region of pronouncedly prolate traps
($\beta<0.5$).
For oblate traps with $\beta=5.2$ or larger, the system is always stable,
irrespective of the value of $N^\sixth\varepsilon$.
This figure presents results obtained in the Gaussian approximation, and so do
all other figures.}
\end{center}
\end{figure}

Dipolar interactions are partially attractive and partially
repulsive, depending on the configuration of the dipoles. One should
keep in mind the simple situation of two dipoles in the plane
perpendicular to their polarization, which repel each other,  as
opposed to the situation of two attracting dipolar particles placed
along the direction of their polarization. Extending this picture to
a cloud of trapped dipoles one would expect that
attraction dominates in prolate traps, and repulsion in oblate
traps (provided the dipoles are polarized along the trap axis as
is the case here). In the case of predominant attraction
one may surmise that instabilities occur.
By varying the two system parameters ($N^\sixth\varepsilon$ and $\beta$)
we have investigated the issue of stability, see Fig.~\ref{fig:stability}.

From this stability diagram one concludes that, for oblate
traps ($\beta>1$), the bigger the trap aspect ratio, the bigger
values of the dipole parameter $N^\sixth\varepsilon$ can be stabilized. In
fact, we
found numerically that fermions form stable configurations
for $\beta>5.2$ irrespective of the strength of their
dipole interaction (an analogous effect was observed for dipolar
bosons by Santos {\it et al.\/} \cite{Santos}). On the other hand,
one might naively expect an exactly opposite effect in
prolate traps ($\beta<1$) where attractive interactions should dominate.
This is not quite true -- indeed for moderate trap
aspect ratios ($1>\beta>0.5$) the critical value of the dipole
parameter is smaller, but in traps that are very soft ($\beta<0.5$)
in the $z$ direction of rotational symmetry we observe an increase of
the critical value of $N^\sixth\varepsilon$ (see the inset in
Fig.~\ref{fig:stability}). This can be understood with the help of the
following argument. We note that the dipole-dipole energy term
vanishes for a uniform density distribution. As the trap is made
softer in the polarization direction, the shape of the cloud along
the soft axis becomes more and more uniform, contributing to the
interaction energy to a lesser extent (this argument was also used
to interpret our earlier results for bosons interacting via
contact and dipole-dipole forces, see \cite{Goral}).

\begin{figure}[!t]
\begin{center}
\begin{picture}(240,325)
\put(-15,5){\epsfig{file=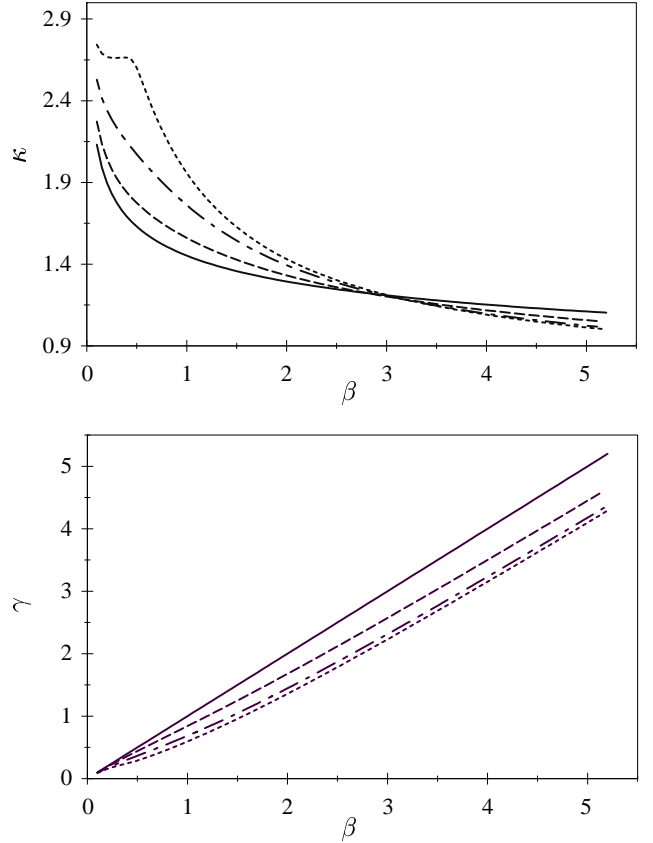}}
\end{picture}
\caption{\label{fig:CloudParams}%
Dependence of cloud parameters $\kappa$ (cloud size; top) and $\gamma$ (cloud
shape; bottom) on $\beta$, the aspect ratio of the trap.
The lines refer to different values of the interaction strength:
$N^\sixth\varepsilon=0$ (solid lines),
$N^\sixth\varepsilon=1$ (dashed lines),
$N^\sixth\varepsilon=2$ (dash-dotted lines),
$N^\sixth\varepsilon=2.4$ (dotted lines).}
\end{center}
\end{figure}

The dependence of $\gamma$ and $\kappa$ on $\beta$, shown in
Fig.~\ref{fig:CloudParams}, is consistent with this argument. We
see that $\gamma$ decreases with decreasing $\beta$ whereas
$\kappa$ increases. Accordingly, the cloud gets stretched along
the $z$ axis of symmetry, and the diameter of the circular cross
section in the $x,y$ plane ($\propto\kappa^{-1}$) is reduced. At
the center of the cloud, we thus have a relatively large volume of
(almost) constant density, and the inhomogeneous parts of the
cloud are relatively far apart. Taken together, these geometric
features lead to a rather small dipole-dipole interaction energy.

\begin{figure}[!t]
\begin{center}
\begin{picture}(230,160)
\put(-3,3){\epsfig{file=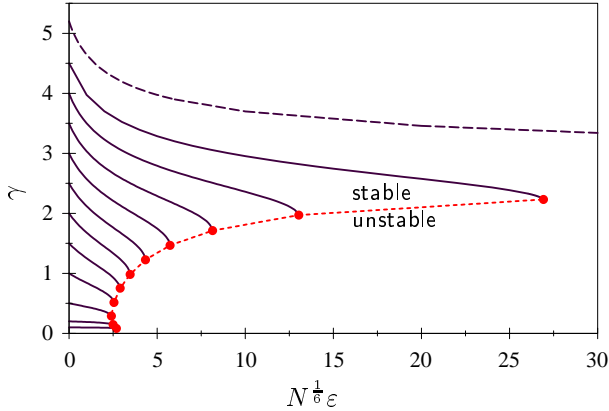}}
\end{picture}
\caption{\label{fig:GammaOfD}%
Aspect ratio $\gamma$ of the cloud as a function of the interaction strength
$N^\sixth\varepsilon$, for various values of the aspect ratio $\beta$ of the
trap.
Note that $\gamma=\beta$ for $N^\sixth\varepsilon=0$.
The dashed line is for $\beta=5.2$, for which the cloud is stable for all
values of $N^\sixth\varepsilon$.}
\end{center}
\end{figure}

Whereas the stability of the system considered can be well
understood, the spatial behavior of the fermionic cloud,
especially near the collapse, seems to be much less intuitive. As we
approach the critical parameter values, the aspect ratio $\gamma$ of the
cloud decreases and the cloud becomes elongated in the attractive
$z$ direction. This type of behavior
is general in the sense that it does not depend on the trap aspect
ratio, see Fig.~\ref{fig:GammaOfD}. It is only
for traps with $\beta>5.2$ that $\gamma$ reaches an asymptotic
value (dependent on $\beta$) for extremely large values of
$N^\sixth\varepsilon$.

\begin{figure}[!b]
\begin{center}
\begin{picture}(230,160)
\put(0,7){\epsfig{file=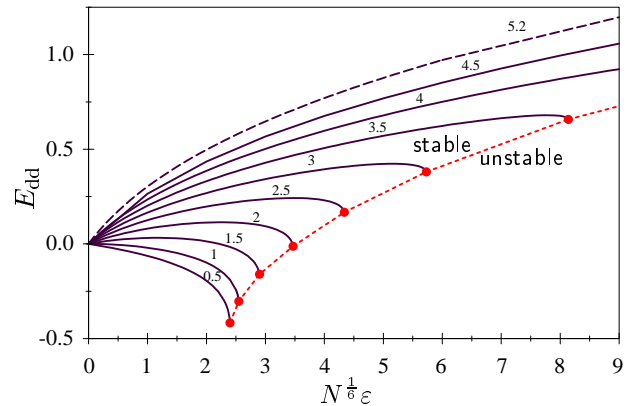}}
\end{picture}
\caption{\label{fig:EddOfD}%
Dipole-dipole interaction energy $\Edd$ as a function of the
interaction strength $N^\sixth\varepsilon$.
The solid lines are for trap aspect ratios $\beta=0.5,1,1.5,\ldots,4.5$;
the dashed line is for $\beta=5.2$.}
\end{center}
\end{figure}

The dependence of the dipolar energy $\Edd$ on the dipole parameter
$N^\sixth\varepsilon$ and the trap geometry is also of interest as it is the
quantity responsible for the (in)stability of the system, see
Fig.~\ref{fig:EddOfD}.
For all prolate traps, $\Edd$ remains negative approaching
some critical value at the collapse point. For $\beta<5.2$,
the dipolar energy can be positive for moderate dipole parameters,
but if their values are large enough, $\Edd$ turns negative indicating
the collapse. For $\beta>5.2$, $\Edd$ is always positive
and increases as a function of $N^\sixth\varepsilon$.

\begin{figure}[!t]
\begin{center}
\begin{picture}(240,170)
\put(0,7){\epsfig{file=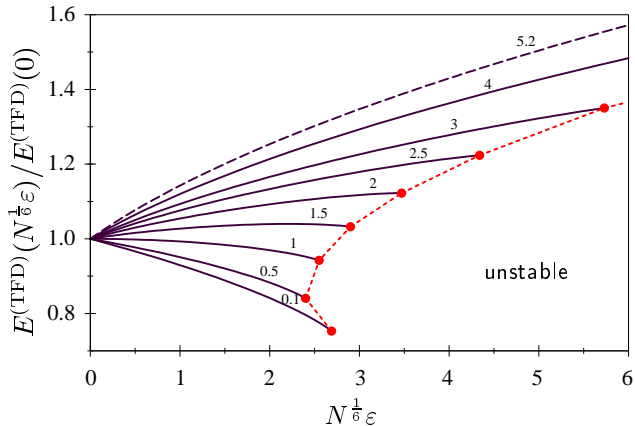}}
\end{picture}
\caption{\label{fig:ETFDofD}%
Normalized TFD energy as a function of the interaction strength for various
trap shapes.
The universal function $e(N^\sixth\varepsilon)$ of (\ref{eq:scale8a}),
normalized to its initial value $e_0=e(0)$, is shown for trap aspect ratios
$\beta=0.1,0.5,1,1.5,\ldots,4$ (solid lines) and $\beta=5.2$ (dashed line).
Note the horizontal slope of the $\beta=1$ line (spherical trap), as required
by (\ref{eq:scale9}).}
\end{center}
\end{figure}

Finally, we take a look at the total TFD energy, see Fig.~\ref{fig:ETFDofD}.
For $\beta<5.2$, the system becomes unstable at the critical value of
$N^\sixth\varepsilon$.
Consistent with Fig.~\ref{fig:stability}, we observe that larger values of
$N^\sixth\varepsilon$ are supported for $\beta=0.1$ and $\beta=1$ than for
$\beta=0.5$.

Let us now discuss the dipole parameter $N^\sixth\varepsilon$ and give
some typical values for it.
Owing to the $N$ dependence, one can locate the experimental system in various
regions of the stability diagram of Fig.~\ref{fig:stability} not only by
choosing (or inducing, as proposed for bosons
\cite{Marinescu,Santos}) a specific value of $\mu$, but also, to some extent,
by varying $N$.
One could also exploit the $\omega$ dependence of
$\varepsilon\propto\sqrt{\omega}$, which is particularly relevant for
optical traps with tight confinement \cite{Haensch+al}.

For the parameters of the fermionic
chromium isotope, $\omega=300$ Hz and $N=10^6$, one obtains
$N^\sixth\varepsilon=0.012$, which is a very small value.
Therefore, the conclusion for atoms possessing even relatively
large magnetic dipole moments is that irrespective of their number
and the trap frequency they will always remain stable against collapse.
The following amusing
analogy offers a good reason for this observation \cite{Rokhsar}.
Let us compare the characteristic sizes of the noninteracting
Fermi gas and the Bose-condensed atomic gas interacting via a
repulsive contact potential. In the Thomas-Fermi approximation the
appropriate quantities, in units of
$\sqrt{\hbar/(M\omega)}$, read: $R_{\rm Fermi}=(48N)^\sixth$ for
fermions and $R_{\rm Bose}=(15Na_{0})^\fifth$ for bosons , where
$a_{0}$ is the scattering length. By equating the two sizes one
can calculate the effective, $N$ dependent, scattering length due to
the exclusion-principle--induced repulsion:
$a_{0}\approx1.68N^{-\sixth}$. For typical numbers of atoms,
$N=10^3\cdots10^6$, this $a_{0}$ is huge on the scale set by typical
scattering lengths for bosonic atoms ($a_0\approx10^{-3}$).
Once we realize how strong is
the repulsion that originates in the Fermi statistics, we understand that
small atomic magnetic dipoles can hardly have a noticeable effect on the
behavior of dipolar fermionic gases. However, for polar molecules
the situation is different. For a trap frequency of $\omega=300$\,Hz and
$N=10^6$ molecules of mass $m\approx100$ a.m.u.
(the typical mass of an alkaline dimer) and a
typical electric dipole moment of 1 Debye, one reaches
$N^\sixth\varepsilon\approx11.5$, which may very well put the system into the
unstable regime -- see Fig.~\ref{fig:stability}.
In this situation, a sufficiently large $\beta$ value will stabilize the system.

In order to assess the quality of our variational results we have
computed exact numerical solutions of Eq.\
(\ref{eq:NonLinIntEq'}). This equation was solved for the density
distribution $g(\svec{x})$ self-consistently starting from the
known analytical result (\ref{eq:n0}) for a non-interacting
($\varepsilon=0$) Fermi gas in a trap \cite{Rokhsar} and slowly
increasing the dipole parameter $N^\sixth\varepsilon$. For each
value of $N^\sixth\varepsilon$ the solution was iterated until
convergence was reached. Then, the value of the dipole parameter
was slightly increased. In order to compute the dipole (integral)
term we note that it has the form of a convolution. Thus, it can
be conveniently evaluated in the Fourier space where it is a
simple product of the Fourier transforms of the density (computed
numerically with the aid of an FFT) and the interaction potential,
the latter being known analytically \cite{Goral}:
\begin{equation}
\int(\D\svec{x})\,\Exp{\I\svec{q}\cdot\svec{x}}
\frac{1-3\cos^2\theta}{|\svec{x}|^3}
=-\frac{4\pi}{3}(1 -3\cos^{2} \alpha) \; ,
\end{equation}
where $\alpha$ is the angle between the Fourier variable $\vec{q}$
and the $z$ direction. In order to assure that the integral term is
evaluated accurately we used a Gaussian distribution for comparison and
chose the grid parameters accordingly.

Our numerical calculations, performed in three dimensions, were quite
demanding so we limited their use to a check of the main features
of the stability diagram in Fig.~\ref{fig:stability}. The solutions obtained
satisfy the virial relations (\ref{eq:scale5}) very well, and
total energies obtained numerically are always below
the corresponding values from the variational analysis. The critical
value of the dipole parameter for a spherical trap is
$N^\sixth\varepsilon=1.96$ as compared to $2.55$ obtained
by the variational calculation.
We also confirmed the effect of increase of the critical
interaction strength that we found, in the Gaussian approximation,
for prolate traps: for $\beta=0.07$ the critical value of $N^\sixth\varepsilon$
is ${2.75(>1.96)}$ as compared to ${2.81(>2.55)}$
obtained analytically in the Gaussian approximation.

Now two remarks about possible extensions of our work are in
order. Firstly, the results presented in this paper describe the
situation of a dipolar fermionic gas at the temperature $T=0$. An
interesting subject of study would be extension of our theory to
finite temperatures. Secondly, there exists a parallel approach in
the Thomas-Fermi model, namely the one in the momentum space
\cite{BGE92,MC+BGE93}. As many experiments with cold gases yield
their momental characteristics, investigation of this alternative
approach also presents an attractive theoretical task.

\acknowledgements
B.-G.~E.\ would like to thank W. Schleich for his support in Ulm where part of
this work was done.
B.-G.~E.\ is grateful for the kind hospitality extended to him in Warsaw.
K.~G.\ acknowledges support by Polish KBN grant no 2~P03B 057 15.
K.~R.\ and K.~G.\ are supported by the subsidy of the Foundation for
Polish Science. Part of the results has been obtained using
computers at the Interdisciplinary Centre for Mathematical and
Computational Modeling (ICM) at Warsaw University.


\begin{references}
\bibitem{comment}
An analogous argument can be used for electric dipoles -- in this case
the contact term is needed to have the curl of an electric field vanish.
\end{references}
\end{document}